\documentclass[journal]{IEEEtran}
\usepackage[export]{adjustbox}
\usepackage{multirow}

\usepackage{xcolor}
\usepackage{lineno,hyperref}
\modulolinenumbers[5]
\usepackage{bm}
\usepackage{comment}
\usepackage{changepage}
\usepackage{amsmath}
\usepackage{graphicx}
\usepackage{subfigure}
\usepackage{multirow}
\usepackage{array}
\newcolumntype{P}[1]{>{\centering\arraybackslash}p{ 1}}
\usepackage{makecell}
\usepackage{pgfplots}
\usepackage{dingbat}
\usepackage[flushleft]{threeparttable}
\usepackage{multirow}
\usepackage{cite}
\definecolor{darkgreen}{rgb}{0.0, 0.5, 0.0}
\newcommand{\diego}[2]{\textcolor{blue}{ 2}}
\begin{document}

%
\title{
QuickQuakeBuildings: Post-earthquake SAR-Optical Dataset for Quick Damaged-building Detection
}

%

\author{Yao~Sun,
        Yi~Wang, 
        and~Michael~Eineder,~\IEEEmembership{Fellow,~IEEE}
\thanks{Y. Sun and Y. Wang are with Data Science in Earth Observation, Technical University of Munich, 80333 Munich, Germany, Y. Wang and M. Eineder are with the Remote Sensing Technology Institute, German Aerospace Center, 82234 Wessling, Germany. (e-mails: yao.sun@tum.de; yi.wang@dlr.de; michael.eineder@dlr.de)
}
\thanks{
The work of Y. Wang is supported by Helmholtz Association through the Framework of Helmholtz AI.}}

\newpage
\thispagestyle{empty}
\onecolumn
\noindent This work has been submitted to the IEEE for possible publication. Copyright may be transferred without notice, after which this version may no longer be accessible.
\newpage
\twocolumn

\maketitle

\begin{abstract}
\textcolor{blue}{This is the preprint version. }
Quick and automated earthquake-damaged building detection from post-event satellite imagery is crucial, yet it is challenging due to the scarcity of training data required for developing robust algorithms. 
This letter presents the first dataset dedicated to detecting earthquake-damaged buildings from post-event very high resolution (VHR) Synthetic Aperture Radar (SAR) and optical imagery. 
{Utilizing open satellite imagery and annotations acquired after the 2023 Turkey–Syria earthquakes, we deliver a dataset of co-registered building footprints and satellite image patches of both SAR and optical data, encompassing more than four thousand buildings.}
The task of damaged building detection is formulated as a binary image classification problem, that can also be treated as an anomaly detection problem due to extreme class imbalance. 
We provide baseline methods and results to serve as references for comparison. 
{Researchers can utilize this dataset to expedite algorithm development, facilitating the rapid detection of damaged buildings in response to future events.}
The dataset and codes together with detailed explanations {and visualization} will be made publicly available at \url{https://github.com/ya0-sun/PostEQ-SARopt-BuildingDamage}.
\end{abstract}


\IEEEpeerreviewmaketitle
\vspace{-0.5cm}

\section{Introduction}

Earthquakes can result in substantial structural and infrastructural damage, often with significant socioeconomic consequences. After an event, 
fast and accurate detection of earthquake-damaged buildings in remote sensing imagery is of great importance. 
Remote sensing technologies can effectively improve the efficiency of disaster management and have been employed to estimate the extent of earthquake damage to buildings~\cite{dell2012remote, dong2013comprehensive, contreras2021earthquake}. 
{Very high resolution (VHR) optical images are easier to interpret, making them a preferred choice for many studies. However, acquisition of cloud-free optical images depends on weather conditions and often needs to wait. 
In contrast, SAR imagery is particularly suitable for rapid disaster response scenarios as it can be acquired regardless of cloud coverage and sun illumination conditions.} Moreover, the enhanced resolution of contemporary SAR satellite images enables the extraction of information at the individual building level~\cite{sun2020cgnet, chen2021cvcmff, sun2021bbox}, comparable to VHR optical data. 

Over the past years, many researchers have developed algorithms utilizing SAR data to detect earthquake-damaged buildings. 
Most existing works detect changes using both pre- and post-earthquake SAR images to acquire building damage information~\cite{miura2016building, guida2010monitoring}; 
however, pre-event high-resolution SAR imagery is generally unavailable in most locations. 
{A few works simulate a SAR image using building shapes extracted from a pre-event optical image and acquisition parameters of a post-event SAR image and detect earthquake-damaged buildings by comparing the simulated and real SAR images ~\cite{brunner2010Earthquake}. }
Utilizing only a single post-event SAR image, some researchers detect damaged buildings by analyzing signatures of destroyed buildings in high-resolution SAR data~\cite{balz2010building,gong2016earthquake,kuny2013signature}. 
{However, the study areas often comprise a limited number of isolated buildings that do not depict the typical conditions in densely populated urban areas, where partially occluded buildings and geometric distortions in SAR images, i.e., foreshortening, layover, and shadowing, commonly exist.} 
Therefore, 
the question of whether a single VHR SAR image acquired after an event can effectively identify damaged buildings remains to be addressed. 
In this regard, 
benchmark datasets play a pivotal role in the development and comparative assessment of diverse methodologies aimed at addressing the following questions: 
To what extent can a single VHR SAR image, acquired post-event, allow to identify damaged buildings and with which accuracy? 
Additionally, how do the outcomes derived from a single post-event SAR image compare with those of an optical image? 

From a practical standpoint, there are several challenges in creating such a dataset: 
1) Limited availability of high-resolution SAR images in disaster-affected areas; 
2) Absence of labels for damaged buildings; 
3) Lack of accurate terrain models, without which aligning the two is a complex task. 
Currently, such a dataset does not exist in the remote sensing field.

{This work presents the first 
dataset for detecting earthquake-damaged buildings in post-event SAR and optical satellite imagery. 
We integrate publicly accessible satellite imagery and annotations obtained following the 2023 Turkey-Syria earthquakes and construct a dataset comprising over four 
thousand buildings, each with satellite image patches of both post-event SAR and optical data and its footprint co-registered with the corresponding image patches. }
We formulate the problem of damaged building detection as an image classification task and benchmark a set of popular machine learning and deep learning methods as a baseline reference. 

The remaining part of the letter proceeds as follows: 
Section \ref{sec:method} introduces the dataset generation approaches,  
Section \ref{sec:Experiments} presents the baseline methods and results, and Section \ref{sec:Conclusion} concludes the paper.

\section{Dataset Generation}\label{sec:method} 

This work aims at addressing building-level damages. The tasks of identifying pixel-level damages and classifying the types of damages are beyond the scope of this dataset. %

Considering the increasing availability of building footprint information across various geographic locations, {we integrate pre-event building footprints with post-event satellite images and assess whether buildings at those locations are damaged.}
The problem of detecting damaged buildings in post-event imagery is therefore formulated as an image classification task with two classes: damaged and intact buildings. 
{We generate and deliver a dataset comprising post-event VHR SAR and optical image patches for each building, building footprint masks corresponding to the image patches, and labels indicating whether each building is damaged or intact. 
The image patches and masks serve as inputs for the algorithm, while the labels represent the ground truth. }

\subsection{Study Area and Data Sources}

The study area is chosen in the city of Islahiye, located in southeastern Turkey near the northwestern border of Syria. 
On February 6, 2023, a magnitude 7.8 earthquake struck Kahramanmaras, Turky, followed by a 7.5 magnitude aftershock nine hours later. The earthquakes inflicted widespread destruction, leading to significant damage to buildings, injuries, and loss of life. 
Islahiye was one of the most affected areas. 

After the earthquakes, a set of high-resolution satellite data was released {under CC BY 4.0 license}\footnote{\url{https://creativecommons.org/licenses/by/4.0/}} to support rescue operations by commercial remote sensing companies, such as Maxar and Planet Lab for optical data, and Capella Space in the SAR domain. During the humanitarian relief efforts, communities across the globe, such as OSM and UN mappers, organized labeling events and identified and validated a significant number of damaged buildings.

\begin{figure}[!]
        \centering
        \includegraphics[trim=0cm 0cm 0cm 0.8cm, clip=false,width=1\columnwidth]{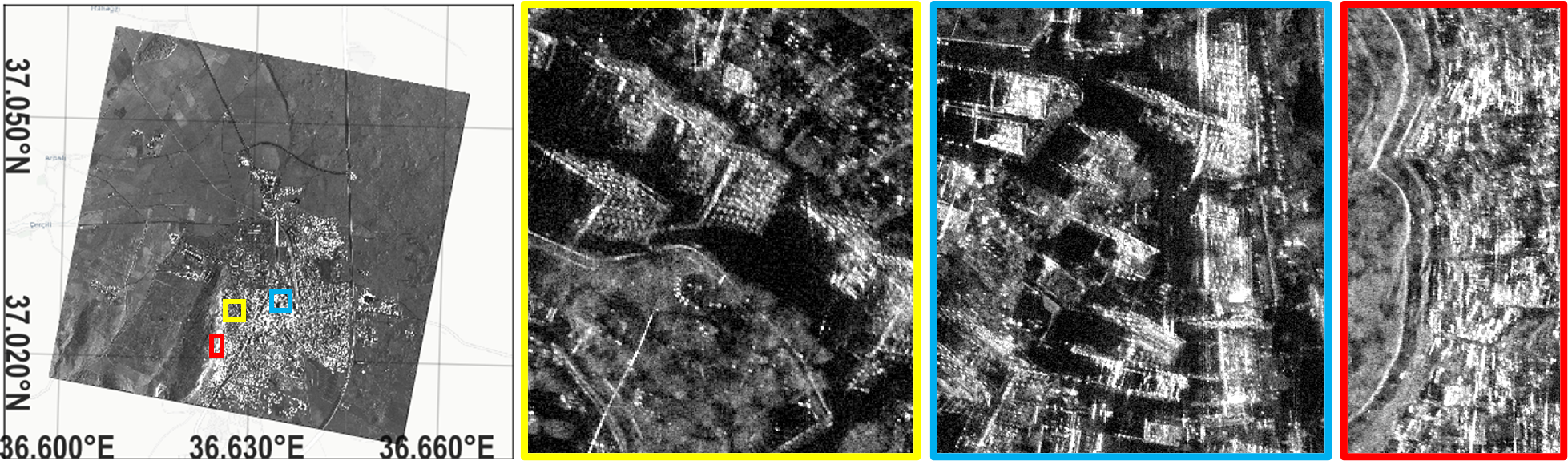}
        \caption{The SAR image coverage and zoomed-in views of three areas in the colored boxes in the SAR image, respectively. }
        \label{fig:sar}
    \end{figure}

We utilize the Spotlight SAR image from Capella Space, acquired on February 9, 2023. The SAR image is of type Geocoded Terrain Corrected (GEO)\footnote{\url{https://support.capellaspace.com/hc/en-us/articles/360039702691-SAR-Data-Formats}}, with a pixel spacing of 0.35 m in both the azimuth and the range direction. The incidence angle of this SAR image is 43.1$^{\circ}$. Figure~\ref{fig:sar} shows the image coverage and zoomed-in views of selected areas. 
The optical image covering the same area was obtained from Maxar Analysis-Ready Data (ARD) under Maxar's open data program. 
The image was acquired on February 7, 2023 by WorldView-3, with a ground sampling distance of 0.31 m and an incidence angle of 83.1$^{\circ}$. 
In addition, we obtained post-event building footprints and labels of destroyed buildings in the study area from Humanitarian OpenStreetMap Team \footnote{\url{https://data.humdata.org/dataset/hotosm\_tur\_buildings}} \footnote{\url{https://data.humdata.org/dataset/hotosm\_tur\_destroyed\_buildings}}. 
Since we consider both SAR and optical images, all used data are chosen or projected to the Universal Transverse Mercator (UTM) coordinate system so that they can be processed uniformly. 
The SAR image is {logarithmically} scaled in dB for further processing.

    \newlength{\tempdima}
    \newcommand{\rowname}[1]
    {\rotatebox{90}{\makebox[\tempdima][c]{\textbf{#1}}}}
    \vspace{-0.5cm}
    \begin{figure}[!]
    \centering
        \rowname{\hspace{2cm}\scriptsize{before registration}} 
        \subfigure{\includegraphics[height=0.238\columnwidth, width=0.238\columnwidth]{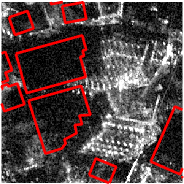}}\hfill  
        \subfigure{\includegraphics[height=0.238\columnwidth, width=0.238\columnwidth]{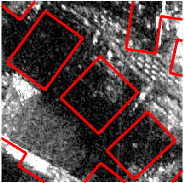}}\hfill
        \subfigure{\includegraphics[height=0.238\columnwidth, width=0.238\columnwidth]{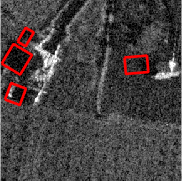}}\hfill
        \subfigure{\includegraphics[height=0.238\columnwidth, width=0.238\columnwidth]{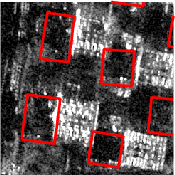}}
\addtocounter{subfigure}{-4} 
        \rowname{\hspace{2cm}\scriptsize{after registration}}
        \subfigure{\includegraphics[height=0.238\columnwidth, width=0.238\columnwidth]{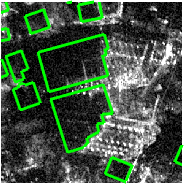}}\hfill
        \subfigure{\includegraphics[height=0.238\columnwidth, width=0.238\columnwidth]{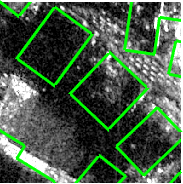}}\hfill
        \subfigure{\includegraphics[height=0.238\columnwidth, width=0.238\columnwidth]{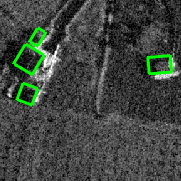}}\hfill
        \subfigure{\includegraphics[height=0.238\columnwidth, width=0.238\columnwidth]{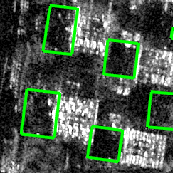}}
    \caption{Examples of four study areas in the SAR image. Building footprint polygons before and after registration are plotted in \textcolor{red}{red} and \textcolor{green}{green}, respectively. }
    \label{fig:ba_reg}
    \end{figure}

\vspace{0.1cm}
\subsection{Co-registration of building footprints and {satellite imagery}}

{Building-level analysis requires accurate registration of 2-D building footprints with satellite images. 
The ARD optical image aligns well with building footprints, requiring no additional registration. 
For the GEO SAR image, 
inspection shows that building polygons are not well-matched with the SAR image, as shown in the first row of Figure~\ref{fig:ba_reg}, and further registration is needed. }

\begin{figure}[!]
    \centering
    \subfigure[]{\includegraphics[width=0.42\columnwidth]{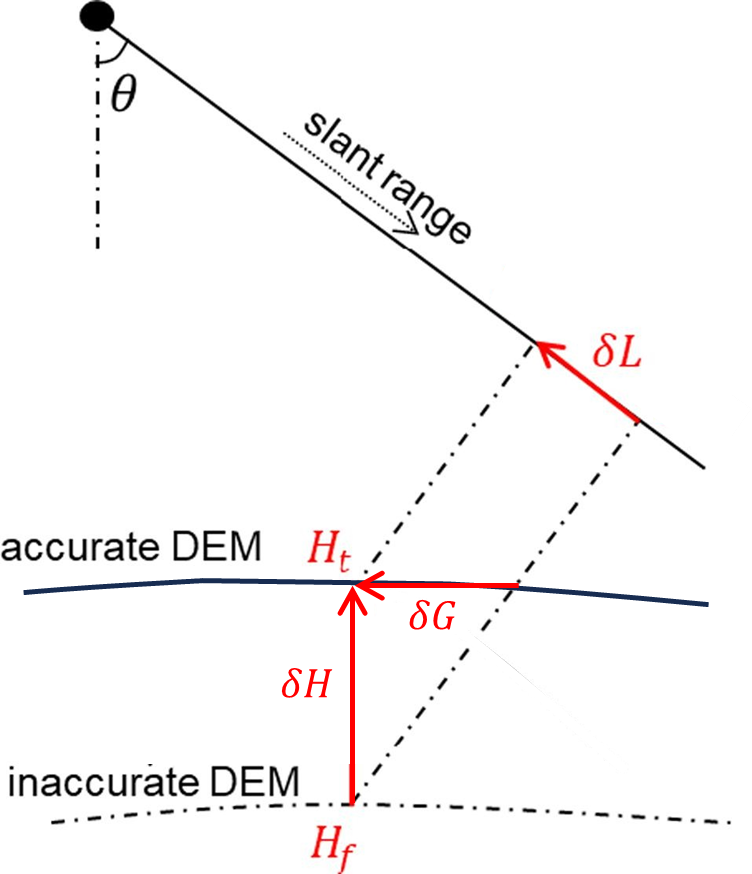}}\hspace{0.7cm}
    \subfigure[]{\includegraphics[width=0.35\columnwidth]{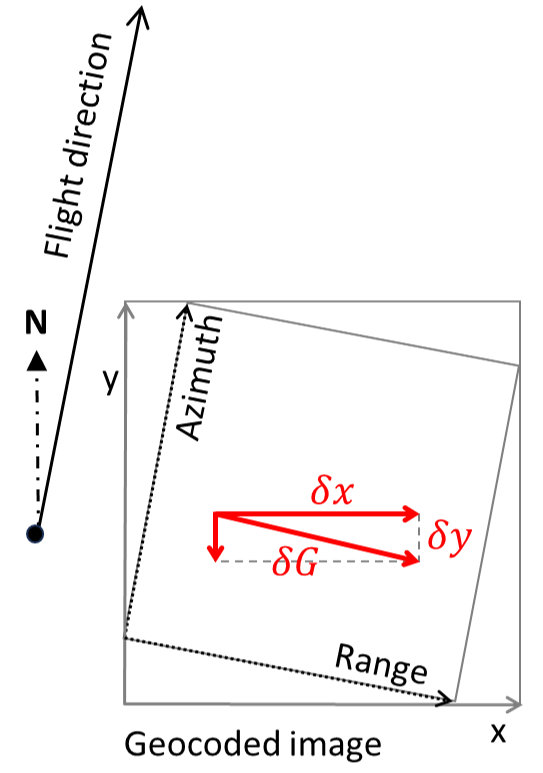}}
    \caption{The geocoding error from inaccurate height. 
    (a) $H_t$ and $H_f$ are the accurate height and inaccurate height of a point, and $\theta$ is the incidence angle. 
    The height error $\delta H$ results in an error of $\delta L = \delta H  cos \theta$ in the slant range and $\delta G = \delta H cot\theta$ on the ground. 
    (b) In the geocoded image, $\delta G$ is decomposed to $\delta x$ and $\delta y$ in the image coordinate system.}
    \label{fig:herror}
\end{figure}

In urban areas, the geocoding errors in SAR data are often caused by inaccurate terrain heights, as illustrated in Figure \ref{fig:herror}(a). 
A height error $\delta H$ causes an error of $\delta L$ in the slant range and {a shift of} $\delta G$ on the ground. For the used SAR image, the incidence angle is 43.1$^{\circ}$; thus, a height error of 10 meters results in an error of 10.69 meters on the ground, causing errors of $\delta x$ and $\delta y$ in the geocoded image related to the flight direction, as shown in Figure \ref{fig:herror}(b). 
The height error $\delta H$ is usually inconstant over the observed area by the SAR sensor; hence, so are the geocoding errors.

\begin{figure}[!]
        \centering
        \includegraphics[trim=0cm 0cm 0cm 0.8cm, clip=true,width=.58\columnwidth]{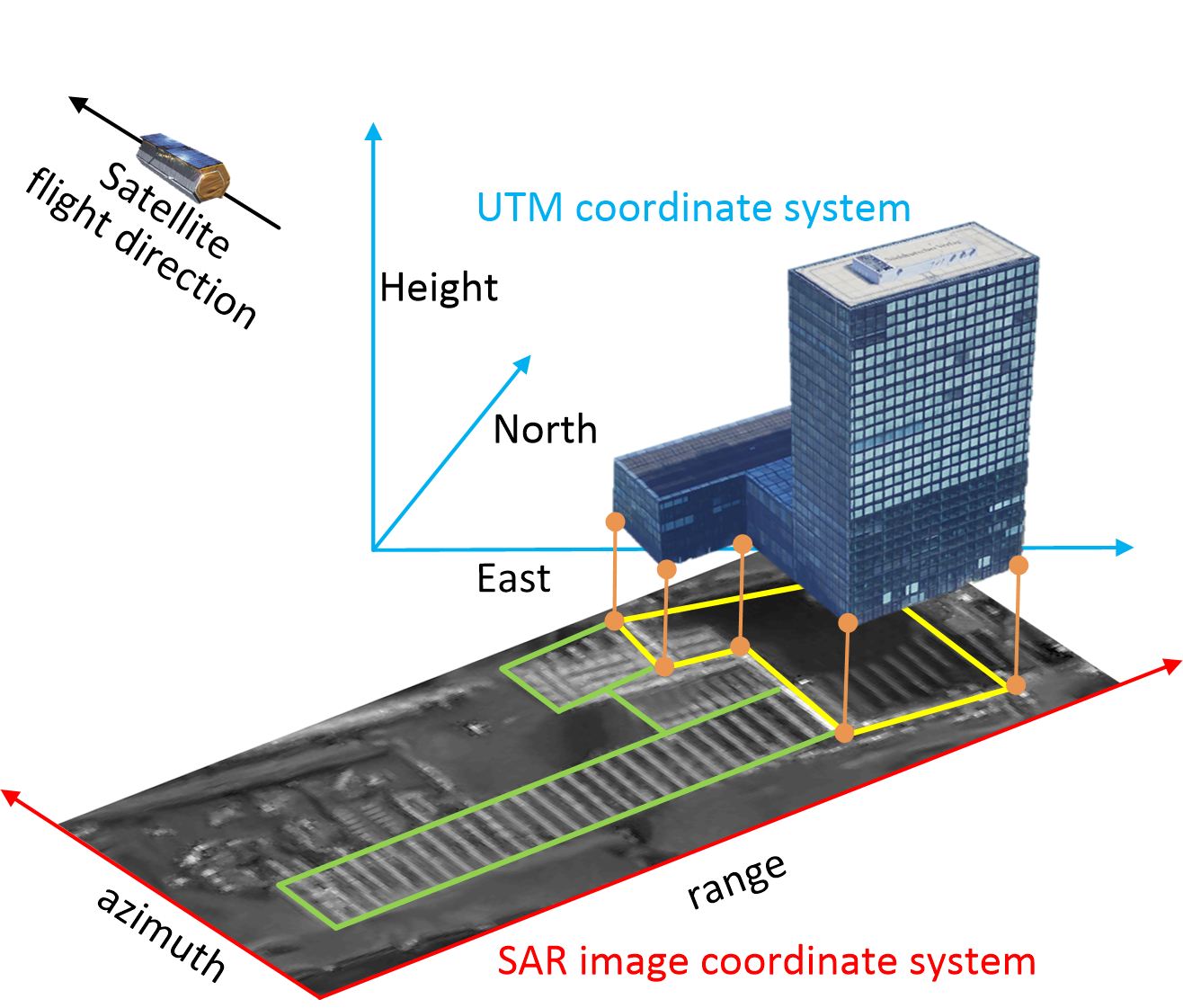}
        \caption{Building signature in SAR imagery and building's geometric correspondence between SAR and building polygon: the near-range side of \textcolor{yellow}{\textbf{the building footprint}} corresponds to the double bounce line in the SAR image, which is \textcolor{darkgreen}{the far-range side of the facade signatures}~\cite{sun2020auto}. }
        \label{fig:cor}
    \end{figure}

To improve the alignment of building polygons and the SAR image, we apply the algorithm developed in~\cite{sun2020auto}, which relies on the corresponding building features representing the bottom of sensor-visible walls in both the two data, i.e., double bounce lines in the SAR image and near-range boundaries of 2-D building polygons, as illustrated in Figure \ref{fig:cor}. 
As the majority of buildings remain upright, with expected double bounce line signatures on the SAR image, the algorithm is applicable.
Interested readers are referred to~\cite{sun2020auto} for more details. Next, we briefly explain the main steps of the algorithms.

\subsubsection{Extracting corresponding features} 

    {In the SAR image, double bounce lines correspond to the far-range side of the bright building signature. The SAR image is first segmented using Potts model~\cite{storath2015joint}. Then, an intensity threshold and an area threshold are applied to select building wall segments.  
    Subsequently, the boundary of wall segments is extracted, and the visibility check is performed to extract the double bounce lines caused by the wall segment boundary.} 
    
    In the building polygons, 
    the segments in the near-rage side of building polygons are extracted from each footprint, which in 3-D represent the bottom of illuminated or partially illuminated walls.

\subsubsection{Registering corresponding features} two point sets, GIS points (from building polygons) and SAR points, are then sampled in the extracted features from both data, and the registration problem is reduced to determine the correspondence and the underlying spatial transformation between two point sets. 

Feature registration consists of three progressive steps: global registration, subarea registration, and polygon registration, and the rigid registration in each step is solved with the Iterative Closest Point (ICP) algorithm~\cite{chen1992object}. 
Global registration uses rough height values for an initial alignment of the two data, ensuring that the residual shift falls within a manageable range. 
Then, a set of grids, i.e., subareas, is evenly distributed over the whole region. The distance between one GIS point and its closest SAR point is calculated for all points. 
If the distribution of all distances within one subarea shows a clear center, the $\delta H$ in each grid is considered to be constant, and subarea registration is conducted. 
When the distribution of all distances does not show a clear center, the constant $\delta H$ assumption does not hold, and the registration proceeds to the polygon level, i.e., finding a rigid transformation for each polygon.  

{Since the SAR image, i.e., the GEO product, has been terrain corrected, we neglect global registration and perform subarea and polygon registration.}
{Afterward, manual validation and editing with expert knowledge are conducted to ensure precise registration, relying on identifying double bounce lines and the assumption of no abrupt terrain changes for editing buildings with unclear double bounce lines~\cite{sun2020auto}. For individual buildings, we estimate a maximum registration error of 5-6 pixels (approximately 2 m) that is sufficient for locating individual buildings. Figure~\ref{fig:ba_reg} shows building polygons on the SAR image in exemplary areas before and after registration. As can be seen, the registration procedure effectively aligns the two data sets. Note that buildings in the mountainous area on the west side of the city (e.g., in or near the red box in Figure~\ref{fig:sar}) are excluded due to unclear signals in the SAR image, posing challenges for registration verification. }

\subsection{Patch generation}

\begin{figure}[!t]
  \begin{tabular}{p{0.01cm}cccccc}
   & \hspace{-0.5cm}\scriptsize{a} & \hspace{-0.5cm}\scriptsize{b} & \hspace{-0.5cm}\scriptsize{c} & \hspace{-0.5cm}\scriptsize{d} & \hspace{-0.5cm}\scriptsize{e} & \hspace{-0.5cm}\scriptsize{f}
   \\
   \hspace{-0.3cm}   \rowname{\hspace{-1.5cm}\tiny{SAR}}& \hspace{-0.5cm} 
    \includegraphics[angle=90,origin=c, width=0.155\columnwidth,valign=t]{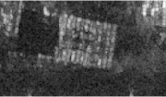}&\hspace{-0.5cm}
    \includegraphics[angle=90,origin=c, width=0.155\columnwidth,valign=t]{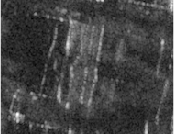}&\hspace{-0.5cm}
    \includegraphics[angle=90,origin=c, width=0.155\columnwidth,valign=t]{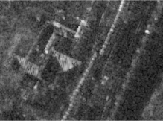}&\hspace{-0.5cm}
    \includegraphics[angle=90,origin=c, width=0.155\columnwidth,valign=t]{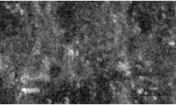}&\hspace{-0.5cm}
    \includegraphics[angle=90,origin=c, width=0.155\columnwidth,valign=t]{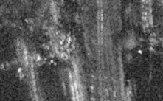}&\hspace{-0.5cm}
    \includegraphics[angle=90,origin=c, width=0.155\columnwidth,valign=t]{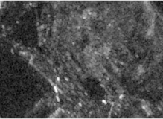}
\\
\\[-1em]
\hspace{-0.3cm}   \rowname{\hspace{-1.5cm}\tiny{SAR footprint}}& \hspace{-0.5cm} 
    \includegraphics[angle=90,origin=c, width=0.155\columnwidth,valign=t]{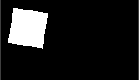}&\hspace{-0.5cm}
    \includegraphics[angle=90,origin=c, width=0.155\columnwidth,valign=t]{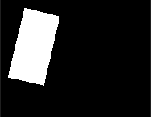}&\hspace{-0.5cm}
    \includegraphics[angle=90,origin=c, width=0.155\columnwidth,valign=t]{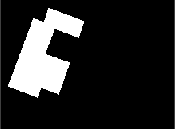}&\hspace{-0.5cm}
    \includegraphics[angle=90,origin=c, width=0.155\columnwidth,valign=t]{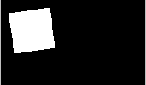}&\hspace{-0.5cm}
    \includegraphics[angle=90,origin=c, width=0.155\columnwidth,valign=t]{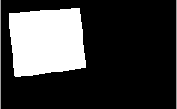}&\hspace{-0.5cm}
    \includegraphics[angle=90,origin=c, width=0.155\columnwidth,valign=t]{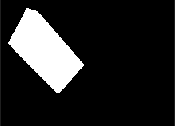}
\\
\\[-1em]
 \hspace{-0.3cm}   \rowname{\hspace{-0.8cm}\tiny{optical}}&\hspace{-0.5cm} 
   \includegraphics[angle=90,origin=c, width=0.155\columnwidth,valign=t]{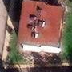}&\hspace{-0.5cm}
    \includegraphics[angle=90,origin=c, width=0.155\columnwidth,valign=t]{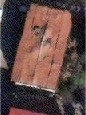}&\hspace{-0.5cm}
    \includegraphics[angle=90,origin=c, width=0.155\columnwidth,valign=t]{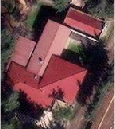}&\hspace{-0.5cm}
    \includegraphics[angle=90,origin=c, width=0.155\columnwidth,valign=t]{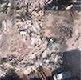}&\hspace{-0.5cm}
    \includegraphics[angle=90,origin=c, width=0.155\columnwidth,valign=t]{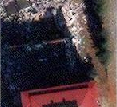}&\hspace{-0.5cm}
    \includegraphics[angle=90,origin=c, width=0.155\columnwidth,valign=t]{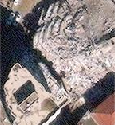}
\\
\\[-1em]
 \hspace{-0.3cm}   \rowname{\hspace{-0.8cm}\tiny{optical footprint}}&\hspace{-0.5cm} 
    \includegraphics[angle=90,origin=c, width=0.155\columnwidth,valign=t]{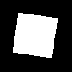}&\hspace{-0.5cm}
    \includegraphics[angle=90,origin=c, width=0.155\columnwidth,valign=t]{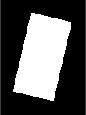}&\hspace{-0.5cm}
    \includegraphics[angle=90,origin=c, width=0.155\columnwidth,valign=t]{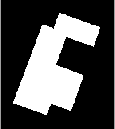}&\hspace{-0.5cm}
    \includegraphics[angle=90,origin=c, width=0.155\columnwidth,valign=t]{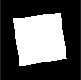}&\hspace{-0.5cm}
    \includegraphics[angle=90,origin=c, width=0.155\columnwidth,valign=t]{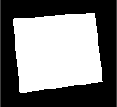}&\hspace{-0.5cm}
    \includegraphics[angle=90,origin=c, width=0.155\columnwidth,valign=t]{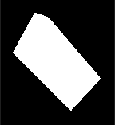}
  \end{tabular}
\caption{Examples of the dataset: a, b, c are \textbf{\textit{intact}} buildings, and d, e, f are \textbf{\textit{damaged}} buildings.
}
\label{fig:eg_patch}
\end{figure}

{For each building, we crop the SAR image and the optical image based on the area of the building, considering including the target building area, i.e., footprint, wall, and roof, as well as the possible ruins around the target building and excluding surrounding buildings.}

{For side-looking SAR images, layover areas of buildings extend from building footprints towards the near-range direction. Therefore, when cropping SAR image patches, a buffer size of 10 pixels, i.e., around 3.5 m, is counted for far-range sides, and for near-range sides, an additional buffer size is counted to include layover areas of the target building in the image patch. 
Specifically, the layover length corresponding to 50 m building height in the ground range direction is decomposed to the image x- and y- directions and added to the buffer size. 
The optical image has a small off-nadir angle of 6.3$^{\circ}$,  resulting in a minor offset between the roof outline and the corresponding 2-D building footprint when the image is not perfectly orthorectified. 
To ensure that the image patches include the entire building roof, we apply a 16-pixel buffer to compensate for the offset, which is approximately 5 meters. This buffer is slightly larger than the one used when cropping the SAR image from the near-range side, and it is used to crop the optical image around the bounding box of the building footprint polygons.}

In addition, for each building, its footprint mask is generated corresponding to the SAR patch and the optical patch, respectively. 
For side-looking SAR data, it is necessary to include the footprint mask to help locate the target building, as the SAR patch may include signals of surrounding buildings. 
For nadir-looking optical data, footprint masks are included for a fair comparison with the SAR data.

Among the 4029 buildings, 169 are damaged, and the remaining 3860 buildings are intact. 
Consequently, the dataset contains 169 damaged buildings and 3860 intact buildings, and each of them has four patches: 
a SAR image patch, a SAR footprint patch, an optical image patch, and an optical footprint patch. Figure~\ref{fig:eg_patch} shows examples of the dataset on six buildings, of which three are intact and three are damaged.

\begin{table*}[ht]
\scriptsize
\setlength{\tabcolsep}{3pt}
\renewcommand{\arraystretch}{1.2}
    \centering
    \caption{{Benchmark results on the dataset: 5-fold mean(std). The highest values of different metrics are highlighted in \textbf{bold} for each set of utilized data. Early Fusion is noted as EF, while Late Fusion is noted as LF.} }
    \label{tab:results}
    \begin{tabular}{l|lllll|lllll|l}
    \hline
        Image & \multicolumn{5}{l|}{\textbf{SAR}} & \multicolumn{5}{l|}{\textbf{optical}} & \textbf{{SAR+optical}} \\
        \hline
        \multirow{2}{*}{model} & \multirow{2}{*}{SVM} & \multirow{2}{*}{RF} & CNN  & ResNet18 & {ResNet18}    & \multirow{2}{*}{SVM} & \multirow{2}{*}{RF} & CNN  & ResNet18  & {ResNet18}  & {ResNet18} \\
        & ~ & ~ &\tiny(EF) &\tiny(EF+ImageNet) & {\tiny(LF+SAR-HUB)} &  &  &\tiny(EF) & \tiny(EF+ImageNet) & {\tiny(LF+ImageNet)} & {\tiny(LF+SAR-HUB+ImageNet)} \\
        \hline
        Precision       & 0.096(0.021)  & 0.223(0.090)  & 0.167(0.028)  & 0.155(0.067)  & \textbf{{0.237(0.116)}}    & 0.163(0.044) & 0.611(0.215) & 0.489(0.154)    & \textbf{{0.666(0.121)}} & {0.622(0.213)} & \textbf{{0.746(0.062)}} \\
        Recall          & \textbf{{0.495(0.202)}} & 0.344(0.208) & 0.371(0.073) & 0.276(0.080) & {0.427(0.104)} & 0.352(0.148) & 0.354(0.098) & 0.432(0.078) & {0.539(0.147)}    & \textbf{{0.646(0.143)}} & \textbf{{0.615(0.117)}} \\
        $F_1$           & 0.154(0.037) & 0.231(0.071) & 0.228(0.032) & 0.184(0.050) & \textbf{{0.282(0.077)}} & 0.207(0.057) & 0.421(0.106) & 0.449(0.096) & {0.581(0.100)}    & \textbf{{0.605(0.120)}} & \textbf{{0.670(0.072)}} \\
        AUROC           & 0.653(0.068) & 0.670(0.079) & 0.739(0.026) & 0.653(0.043) & \textbf{{0.769(0.032)}} & 0.723(0.047) & 0.810(0.062) & 0.853(0.045) & {0.938(0.022)}  & \textbf{{0.941(0.026)}} & \textbf{{0.962(0.023)}} \\
        \hline

    \end{tabular}
\end{table*}

\section{Experimental Results}\label{sec:Experiments} 
We benchmark an image classification task with two classes: damaged and intact buildings. Due to the significant class imbalance, it can also be viewed as an anomaly detection task, i.e., detecting damaged buildings within the entire dataset.

\subsection{Baseline Approaches}

{Four models are introduced to establish a benchmark: 
support vector machine (SVM), 
random forest (RF), 
3-layer convolutional neural network (CNN),
and ResNet-18~\cite{he2016deep}. }

SVM and RF are selected for their good performance in many applications, including classifying collapsed and standing buildings from post-event SAR imagery as reported in~\cite{gong2016earthquake}. 
To extract features as the input, we mask out non-building pixels with footprints, and follow the setup proposed in~\cite{gong2016earthquake}, which employs four first-order statistics, i.e., mean, variance, skewness, and kurtosis, and eight second-order image statistical measures, i.e., mean, variance, homogeneity, contrast, dissimilarity, entropy, second moment, and correlation. 
For a detailed explanation, the readers are referred to \cite{gong2016earthquake}.

{A simple 3-layer CNN and a ResNet-18 are selected as the deep learning backbones. The simple CNN consists of 3 convolution-ReLU-maxpool blocks, followed by average pooling, a linear layer with dropout, and a final classification layer. We stack image and building footprint as the input. The ResNet-18 follows the standard design in~\cite{he2016deep}, with which we benchmark both early and late fusion results of images and building footprints. For late fusion, images and building footprints are encoded by separate encoders, and the feature vectors are concatenated together to a following linear layer with dropout and a final classification layer. Apart from single-modal results, we also conduct a late fusion experiment with both SAR and optical data as a multimodal reference.}

\subsection{Evaluation Metrics}

To evaluate the performance of the baseline methods, we report the precision, recall, and $F_1$ scores: 

\begin{equation}
    P = \frac{tp}{tp + fp}, 
    R = \frac{tp}{tp + fn},
    F_1 = 2\cdot\frac{P \cdot R}{P + R}, 
\end{equation}
where $P$ and $R$ denote the precision and recall, and $tp$, $fp$, $tn$, $fn$ represent true positives, false positives, true negatives, and false negatives for buildings, respectively. The thresholds for positive/negative are determined by best $F_1$ scores.

In addition, we report the area under the receiver-operator curve (AUROC), 
{a standard metric in anomaly detection tasks that well reflects the model's efficiency in distinguishing between classes}. The AUROC score summarizes the ROC curve into a single number that describes the performance of a model for multiple thresholds at the same time.

\subsection{Implementation Details}

We conduct cross-fold experiments for a robust evaluation on the relatively small dataset. Specifically, we split the dataset into 5 folds with a balanced number of damaged and intact buildings, and run each experiment 5 times with 4 folds training and 1 fold testing. The mean and standard deviation are calculated and reported for every evaluation metric.

We preprocess the images by removing 2\% pixel outliers and normalizing the pixel values to the range [0,1]. 
{For simple CNN, we randomly initialize the model; for ResNet-18 early fusion, we use pure ImageNet weights; for ResNet-18 late fusion, we use ImageNet weights for optical images and footprints, and SAR-HUB \cite{sarhub} weights for SAR images.}
We use random resized crop and random horizontal and vertical flip as data augmentations. To deal with the significant class imbalance, we give a bigger weight to damaged buildings and a smaller weight to intact buildings during data sampling. We optimize binary cross entropy loss with AdamW optimizer for 30 epochs. The learning rate follows a cosine-decay schedule starting from 0.0001. Batch size is set to 32.

For the statistic features used by the two machine learning methods, we calculated the features of image patches based on the implementation of 
PyFeats\footnote{\url{https://github.com/giakou4/pyfeats}}. For calculating the evaluation metrics, we use the implementation in scikit-learn\footnote{\url{https://scikit-learn.org/stable/modules/generated/sklearn.metrics.roc_auc_score.html}}.

\subsection{Performance Comparison} 

Table~\ref{tab:results} illustrates variations in performance across different models on the dataset. 
{In general, deep neural networks outperform SVM and RF in both SAR and optical scenarios.}
{For the SAR image, ResNet-18 underperforms simple CNN regarding all four metrics with early fusion and ImageNet weights, indicating the optimization challenge of complex SAR data. This issue is resolved with proper weight initialization from SAR-HUB (around 10\% improvement in AUROC and $F_1$ scores), highlighting the importance of SAR pretrained models such as provided in~\cite{sarhub}.}

For the optical image, ResNet-18 with late fusion stands out with the highest recall, $F_1$ score, and AUROC. It outperforms other models across these three metrics. ResNet-18 with early fusion gived best precision. 
The other deep learning model, CNN, demonstrates a balance between precision and recall, resulting in a high $F_1$ score and an impressive AUROC of 0.853.
RF attains high precision at 0.611 but a relatively lower recall rate. In contrast, SVM yields less favorable outcomes compared to other models. 

Comparing SAR and optical images, we can see that SAR images appear to be more challenging for all models, with generally lower performance than optical images. {Nevertheless, good SAR models, e.g., ResNet-18 with SAR-HUB pretrained weights, outperform bad optical models such as SVM.} 

{In addition, the fusion of SAR and optical imagery provides further improvement compared to each single modality, confirming the complementary information across different modalities.}

\section{Conclusion}\label{sec:Conclusion} 

Detecting earthquake-damaged buildings in post-event satellite imagery is essential yet challenging. This study introduces a dataset designed to address the issue and to foster the development of robust algorithms. 
The dataset combines post-event SAR and optical satellite images with labels of damaged and intact buildings, and the problem is formulated as an image classification task. 
{We provide a benchmark on both modalities with different baseline methods and a baseline of fusing optical and SAR data.} Results show that detecting damage from post-event SAR images is valuable and possible but more challenging than optical images. Such findings call for further research on improved methods, in particular on SAR images. {In addition, the performance gain through simple SAR-optical fusion verifies the potential in using multimodal data when they can be acquired following an event. }

Constrained by limited data quantity and data imbalance, this dataset serves as a starting point. The dataset will undergo expansion and updates as new data emerges in the future. We hope that the research community will engage in further algorithm development for post-earthquake damaged building assessment in SAR images and, where feasible, share their data, thus expediting the identification of post-disaster damaged structures. 




\ifCLASSOPTIONcaptionsoff
  \newpage
\fi

\bibliographystyle{IEEEtran}
\tiny
\bibliography{reference}

\begin{thebibliography}{10}
\providecommand{\url}[1]{#1}
\csname url@samestyle\endcsname
\providecommand{\newblock}{\relax}
\providecommand{\bibinfo}[2]{#2}
\providecommand{\BIBentrySTDinterwordspacing}{\spaceskip=0pt\relax}
\providecommand{\BIBentryALTinterwordstretchfactor}{4}
\providecommand{\BIBentryALTinterwordspacing}{\spaceskip=\fontdimen2\font plus
\BIBentryALTinterwordstretchfactor\fontdimen3\font minus \fontdimen4\font\relax}
\providecommand{\BIBforeignlanguage}[2]{{%
\expandafter\ifx\csname l@#1\endcsname\relax
\typeout{** WARNING: IEEEtran.bst: No hyphenation pattern has been}%
\typeout{** loaded for the language `#1'. Using the pattern for}%
\typeout{** the default language instead.}%
\else
\language=\csname l@#1\endcsname
\fi
#2}}
\providecommand{\BIBdecl}{\relax}
\BIBdecl

\bibitem{dell2012remote}
F.~Dell'Acqua and P.~Gamba, ``Remote sensing and earthquake damage assessment: Experiences, limits, and perspectives,'' \emph{Proceedings of the IEEE}, vol. 100, no.~10, pp. 2876--2890, 2012.

\bibitem{dong2013comprehensive}
L.~Dong and J.~Shan, ``A comprehensive review of earthquake-induced building damage detection with remote sensing techniques,'' \emph{ISPRS Journal of Photogrammetry and Remote Sensing}, vol.~84, pp. 85--99, 2013.

\bibitem{contreras2021earthquake}
D.~Contreras, S.~Wilkinson, and P.~James, ``Earthquake reconnaissance data sources, a literature review,'' \emph{Earth}, vol.~2, no.~4, pp. 1006--1037, 2021.

\bibitem{sun2020cgnet}
Y.~Sun, Y.~Hua, L.~Mou, and X.~X. Zhu, ``{CG-Net}: {Conditional} {GIS}-aware network for individual building segmentation in {VHR SAR} images,'' \emph{IEEE Transactions on Geoscience and Remote Sensing}, pp. 1--15, 2021.

\bibitem{chen2021cvcmff}
J.~Chen, X.~Qiu, C.~Ding, and Y.~Wu, ``{CVCMFF Net: Complex-valued convolutional and multifeature fusion network for building semantic segmentation of InSAR images},'' \emph{IEEE Transactions on Geoscience and Remote Sensing}, vol.~60, pp. 1--14, 2021.

\bibitem{sun2021bbox}
Y.~Sun, L.~Mou, Y.~Wang, S.~Montazeri, and X.~X. Zhu, ``Large-scale building height retrieval from single sar imagery based on bounding box regression networks,'' \emph{ISPRS Journal of Photogrammetry and Remote Sensing}, vol. 184, pp. 79--95, 2022.

\bibitem{miura2016building}
H.~Miura, S.~Midorikawa, and M.~Matsuoka, ``{Building damage assessment using high-resolution satellite SAR images of the 2010 Haiti earthquake},'' \emph{Earthquake Spectra}, vol.~32, no.~1, pp. 591--610, 2016.

\bibitem{guida2010monitoring}
R.~Guida, A.~Iodice, and D.~Riccio, ``{Monitoring of collapsed built-up areas with high resolution SAR images},'' in \emph{2010 IEEE International Geoscience and Remote Sensing Symposium}.\hskip 1em plus 0.5em minus 0.4em\relax IEEE, 2010, pp. 2422--2425.

\bibitem{brunner2010Earthquake}
D.~Brunner, G.~Lemoine, and L.~Bruzzone, ``Earthquake damage assessment of buildings using {VHR} optical and {SAR} imagery,'' \emph{IEEE Transactions on Geoscience and Remote Sensing}, vol.~48, no.~5, pp. 2403--2420, 2010.

\bibitem{balz2010building}
T.~Balz and M.~Liao, ``{Building-damage detection using post-seismic high-resolution SAR satellite data},'' \emph{International Journal of Remote Sensing}, vol.~31, no.~13, pp. 3369--3391, 2010.

\bibitem{gong2016earthquake}
L.~Gong, C.~Wang, F.~Wu, J.~Zhang, H.~Zhang, and Q.~Li, ``{Earthquake-induced building damage detection with post-event sub-meter VHR TerraSAR-X staring spotlight imagery},'' \emph{Remote Sensing}, vol.~8, no.~11, p. 887, 2016.

\bibitem{kuny2013signature}
S.~Kuny, K.~Schulz, and H.~Hammer, ``{Signature analysis of destroyed buildings in simulated high resolution SAR data},'' in \emph{2013 IEEE International Geoscience and Remote Sensing Symposium-IGARSS}, 2013.

\bibitem{sun2020auto}
Y.~Sun, S.~Montazeri, Y.~Wang, and X.~X. Zhu, ``Automatic registration of a single {SAR} image and {GIS} building footprints in a large-scale urban area,'' \emph{ISPRS Journal of Photogrammetry and Remote Sensing}, vol. 170, pp. 1--14, 2020.

\bibitem{storath2015joint}
M.~Storath, A.~Weinmann, J.~Frikel, and M.~Unser, ``{Joint image reconstruction and segmentation using the Potts model},'' \emph{Inverse Problems}, vol.~31, no.~2, p. 025003, 2015.

\bibitem{chen1992object}
Y.~Chen and G.~Medioni, ``Object modelling by registration of multiple range images,'' \emph{Image and vision computing}, vol.~10, no.~3, pp. 145--155, 1992.

\bibitem{he2016deep}
K.~He, X.~Zhang, S.~Ren, and J.~Sun, ``Deep residual learning for image recognition,'' in \emph{Proceedings of the IEEE conference on computer vision and pattern recognition}, 2016, pp. 770--778.

\bibitem{sarhub}
H.~Yang, X.~Kang, L.~Liu, Y.~Liu, and Z.~Huang, ``{SAR-HUB: Pre-Training, Fine-Tuning, and Explaining},'' \emph{Remote Sensing}, vol.~15, no.~23, 2023.

\end{thebibliography}


\end{document}